\documentclass[aps,prd,notitlepage,onecolumn,showpacs,superscriptaddress,groupedaddress,nofootinbib]{revtex4-1}  
\usepackage{graphicx}  
\usepackage{dcolumn}   
\usepackage{bm}        
\usepackage{bbm}        
\usepackage{amssymb}   
\usepackage{amsmath}
\usepackage{slashed}
\usepackage{mathtools}
\usepackage{hyperref}
\usepackage{cleveref}
\usepackage{multirow}

\allowdisplaybreaks

\usepackage{tikz}
\usetikzlibrary{arrows,shapes}
\usetikzlibrary{trees}
\usetikzlibrary{matrix,arrows} 				
\usetikzlibrary{positioning}				
\usetikzlibrary{calc,through}				
\usetikzlibrary{decorations.pathreplacing}  
\usepackage{pgffor}							

\usetikzlibrary{decorations.pathmorphing}	
\usetikzlibrary{decorations.markings}
\tikzset{
    vector/.style={decorate, decoration={snake}, draw},
	provector/.style={decorate, decoration={snake,amplitude=2.5pt}, draw},
	antivector/.style={decorate, decoration={snake,amplitude=-2.5pt}, draw},
    fermion/.style={draw=black, postaction={decorate},
        decoration={markings,mark=at position .55 with {\arrow[draw=black]{>}}}},
    fermionbar/.style={draw=black, postaction={decorate},
        decoration={markings,mark=at position .55 with {\arrow[draw=black]{<}}}},
    fermionnoarrow/.style={draw=black},
    gluon/.style={decorate, draw=black,
        decoration={coil,amplitude=4pt, segment length=5pt}},
    scalar/.style={dashed,draw=black, postaction={decorate},
        decoration={markings,mark=at position .55 with {\arrow[draw=black]{>}}}},
    scalarbar/.style={dashed,draw=black, postaction={decorate},
        decoration={markings,mark=at position .55 with {\arrow[draw=black]{<}}}},
    scalarnoarrow/.style={dashed,draw=black},
    electron/.style={draw=black, postaction={decorate},
        decoration={markings,mark=at position .55 with {\arrow[draw=black]{>}}}},
	bigvector/.style={decorate, decoration={snake,amplitude=4pt}, draw},
}

\definecolor{Gray}{gray}{0.8}
\definecolor{GrayLight}{gray}{0.9}
\definecolor{Darkgreen}{RGB}{30,120,30}

\tikzstyle{block} = [draw, rectangle, 
    minimum height=3em, minimum width=6em]

\newcommand{\Lagr}{\mathcal{L}}

\newcommand{\nn}{\nonumber \\}

\begin{document}

\hspace{5.2in} \mbox{CALT-TH/2021-014}

\title{ Simple models with both  baryon  and lepton number violation by two units}
\author{Andreas Helset, Clara Murgui and Mark B. Wise}                             


\affiliation{Walter Burke Institute for Theoretical Physics, California Institute of Technology, Pasadena, CA 91125, USA}

\date{\today}

\begin{abstract}
	We construct simple renormalizable extensions of the standard model where the leading baryon number violating processes have $\Delta B= \pm\Delta L =- 2$. These models contain additional scalars. The simplest models contain a color singlet and a colored sextet. For such baryon number violation to be observed in experiments, the scalars cannot be much heavier than a few TeV. We find that such models are strongly constrained by LHC physics, LEP physics, and flavor physics.
\end{abstract}

\maketitle

\section*{Introduction}

If the observed baryon asymmetry of the universe arises from physics below the Planck scale, then it signals new physics that most likely fits into the current paradigm of quantum field theory. Presuming that this is the case, it is interesting to enumerate the possible (non-renormalizable) contact interactions that give rise to these processes and  construct renormalizable extensions of the standard model that produce them. Lorentz invariance and hypercharge invariance restrict standard model operators that change baryon number, $\Delta B$, and possibly lepton number, $\Delta L$, to have $(\Delta B - \Delta L) / 2$ be an integer \cite{Kobach:2016ami}. 
The simplest case is\footnote{In this paper we consider local operators without covariant derivatives that are composed of only fermion fields. For example, a case we are not considering is $\Delta B = - \Delta L = -1$ which occurs at dimension 7 and involves a Higgs field or a covariant derivative.} $\Delta B=\Delta L=-1$; for example, $p \rightarrow e^+ \pi^0$. In this case, grand unified theories provide well-motivated renormalizable models that realize this type of process. At lower energies the effects are represented by dimension-six operators suppressed by a mass scale $\Lambda$ squared, and limits from laboratory experiments searching for proton decay imply that  $\Lambda > 10^{16}~{\rm GeV}$. 

Organizing by increasing mass dimension of the local operators that give rise to the baryon number violating processes, the next case is dimension-9 operators with $\Delta B=-2, ~\Delta L=0$ or $\Delta B =-1, ~\Delta L = -3$. The first case could produce for example neutron-antineutron oscillations. The current experiments limit the mass scale that suppresses such operators at $\Lambda> 500~{\rm TeV}$. All the simplest renormalizable models that give rise to these processes have been constructed \cite{Arnold:2012sd}. 
In both cases the scale $\Lambda$ is so high that there are no relevant constraints on the new degrees of freedom and their couplings to quarks from flavor physics and LHC experiments.

The next cases of interest are dimension-12 operators which lead to $\Delta B=\pm \Delta L= -2$ processes or $\Delta B = -1, ~\Delta L=-5$ processes. In this paper we focus on the first of these where the leading local operators contain six quark fields and two lepton fields. 
Dimensional analysis gives a rate, $\Gamma \sim \Lambda_{\rm QCD}^{15}M_N^2/\Lambda^{16}$, for such low-energy processes. Using $\Lambda_{\rm QCD}=200~{\rm MeV}$ for the non-perturbative strong interaction scale, $M_N \simeq 1~{\rm GeV}$, and $\Lambda= 3~{\rm TeV}$, this crude estimate of the lifetime for $\Delta B= \pm \Delta L = - 2$ processes is $\tau=1/\Gamma \sim 10^{34}$ years. The high power of the non-perturbative scale $\Lambda_{\rm QCD}$ that $\tau$ depends on suggests that model-dependent estimates of the hadronic matrix elements relevant for $\Delta B= \pm \Delta L= - 2$ processes may have a high degree of uncertainty.
 
 For processes with $|\Delta B| >2$ the scale of new physics must be below the weak scale for such processes to be observable in the laboratory. There will not be any renormalizable models that are consistent with experiments and give rise to such processes that are observable in the laboratory.

In this paper, we enumerate the relevant dimension-12 $\Delta B = \pm \Delta L = -2$ operators by
applying Hilbert series techniques. 
Then we discuss simple renormalizable models that give rise to some of these operators after integrating out the heavy scalars. 
Earlier works along these lines occurred in refs.~\cite{He:2021mrt,Gardner:2018azu}.
The simplest models, of which there are two, add two new representations of scalars to the standard model, where one of the scalars is a color sextet. 
For the two simplest models we discuss the phenomenology in detail. The models are strongly constrained by flavor changing neutral current processes and---without making some specific choices for the couplings of the new scalars and assumptions about the right-handed quark mass eigenstate fields---will not permit  measurable baryon number violation in the laboratory. 
Even when the models escape constraints from flavor physics, there are still strong constraints on the masses and couplings of the new scalars from LHC and LEP physics. 
Given the large uncertainty in our estimate of the hadronic matrix element, one of the non-minimal models we consider may give an observable rate for the  process $pp \rightarrow \mu^+\mu^+$ (at the nuclear level $(A,Z)\rightarrow (A-2, Z-2)+\mu^{+} \mu^{+}$).


\section*{Baryon number violating operators}

We are considering simple models
where the leading baryon number violating processes have\footnote{We define $\Delta B$ as the number of baryons in the final state minus the number of baryons in the initial state. $\Delta L$ is defined with the same sign convention.}
$\Delta B = \pm \Delta L = -2$.
The corresponding effective operators have mass dimension 12.
Using Hilbert series techniques---which have been developed for the standard
model \cite{Lehman:2015coa,Lehman:2015via,Henning:2015daa,Henning:2015alf,Henning:2017fpj}---we can enumerate the relevant operators.
We use the standard model fermions fields $\{ Q_{L}, u_{R}, d_{R} , L_{L}, e_{R} \}$, which have the following quantum numbers
\begin{align}
        Q_{L} \sim (\bm{3}, \bm{2})_{1/6}, \qquad
        u_{R} \sim (\bm{3}, \bm{1})_{2/3}, \qquad
        d_{R} \sim (\bm{3}, \bm{1})_{-1/3}, \qquad
        L_{L} \sim (\bm{1}, \bm{2})_{-1/2}, \qquad
        e_{R} \sim (\bm{1}, \bm{1})_{-1} ,
\end{align}
and the Hermitian conjugate fields, i.e., we don't include right-handed neutrinos.
In the description of the standard model quantum numbers the first entry is the color representation, the second is the weak $SU(2)$ representation, and the subscript is the hypercharge. 
The subscripts on the fermions listed above indicate the representations
of the Lorentz group $SU(2)_{L} \otimes SU(2)_{R}$.
The Hilbert series gives the number of independent operators for a given
field content, taking into account redundancies coming from field
redefinitions and integration-by-parts relations.
The dimension-12 $\Delta B = \pm \Delta L = - 2$ operators are listed in table~\ref{tab:operators}. For each operator in table~\ref{tab:operators} we list the models we consider in this paper which give rise to it. Ref.~\cite{He:2021mrt} constructed a complete operator basis with $\Delta B = \Delta L = \pm 2$ for one generation of quarks and estimated some hadronic matrix elements of these operators.
\begin{table}
\begin{center}
 \begin{tabular}{||c  | c | c | c ||} 
 \hline
 Operator & $\#$ operators & $\#$ terms & Model \\ [0.5ex] 
 \hline\hline
 $u d d d d d L^{\dagger} L^{\dagger}$ & $\frac{1}{24} n_{L} n^{3}_{Q} ( n_{Q} + 1) \left[ 2( -1 +  n^{2}_{Q}) +  n_{L} ( 8 - 3 n_{Q} + 7 n^{2}_{Q} ) \right]$ & 2 &  -  \\ [1ex] 
 \hline
 $d d d d Q Q L^{\dagger} L^{\dagger}$ &   $\frac{1}{48} n_{L} n^{2}_{Q} \left[ 6 + 35 n_{Q} + n^{2}_{Q}(6 + n_{Q}) + n_{L} n_{Q} ( 6 + 11 n_{Q} + n^{2}_{Q} (6 + 25 n_{Q})) \right]$ &  2 &  -     \\
 \hline
 $d d d d d Q L^{\dagger} e^{\dagger}$ & $\frac{1}{24} n^{2}_{L} n^{3}_{Q} ( n_{Q} + 1 )\left[ 10 + n_{Q} (-3 + 5 n_{Q}) \right]$ & 1 & -\\
 \hline
 $d d d d d d e^{\dagger} e^{\dagger}$ & $\frac{1}{288} n_{L} (n_{L} + 1) n^{2}_{Q} (n_{Q} + 1) \left[50 + 19 n_{Q} + n^{2}_{Q} (5 n_{Q} - 2)\right]$ & 1  & \hyperref[sec:modelII]{II} , \hyperref[sec:modelVII]{VII} , \hyperref[sec:modelVIII]{VIII}   \\
  \hline
 $u u u u d d e e$ & $\frac{1}{48} n_{L} n^{2}_{Q} \left[ 6 + 31 n_{Q} + 4 n^{2}_{Q} - n^{3}_{Q} (7 + 10 n_{Q}) + n_{L} n_{Q} ( 10 + 13 n_{Q} + 7 n^{2}_{Q}( 2 + 5 n_{Q})) \right]$ & 4 &  \hyperref[sec:modelVI]{VI}   \\
  \hline
 $u u u d d d L L$ & $\frac{1}{72} n_{L} (n_{L} - 1) n^{2}_{Q} \left[ 16 + 12 n_{Q} + 13 n^{2}_{Q}+ n^{3}_{Q} (6 + 25 n_{Q}) \right]$ & 3 & \hyperref[sec:modelI]{I} , \hyperref[sec:modelV]{V}  \\
  \hline
 $u u u d d Q L e$ & $\frac{1}{12} n^{2}_{L} n^{3}_{Q} \left[ 2 + 5 n_{Q} + n^{2}_{Q} (4 + 25 n_{Q}) \right]$ & 4 & -  \\
  \hline
 $u u u d Q Q e e$ & $\frac{1}{24} n_{L} n^{3}_{Q} \left[ - (n_{Q} + 1) (2 + n_{Q} (-1 + 5 n_{Q})) + n_{L} (-2 + 5 n_{Q} + n^{2}_{Q} (8 + 25 n_{Q})) \right]$ & 3 &  \hyperref[sec:modelVI]{VI}  \\
  \hline
 $u u d d Q Q L L$ & $\frac{1}{2} n_{L} n^{3}_{Q} \left[ 3 - n^{2}_{Q} + n_{L} n_{Q} (1 + 5 n^{2}_{Q}) \right]$  & 4 & \hyperref[sec:modelI]{I} ,   \hyperref[sec:modelV]{V}   \\
  \hline
 $u u d Q Q Q L e$ & $\frac{1}{3} n^{2}_{L} n^{4}_{Q} \left[ -1 + 10 n^{2}_{Q} \right]$ & 7 & - \\
  \hline
 $u u Q Q Q Q e e$ & $\frac{1}{12} n_{L} n^{3}_{Q} \left[ 7 - n^{2}_{Q} + n_{L} n_{Q} (1 + 5 n^{2}_{Q})\right]$ & 3  & \hyperref[sec:modelIV]{IV} ,  \hyperref[sec:modelVI]{VI} \\
 \hline
 $u d Q Q Q Q L L$ & $\frac{1}{48} n_{L} n^{3}_{Q} \left[ 6 - 19 n_{Q} - n^{2}_{Q} (6 + 5 n_{Q}) + n_{L}( - 6 - 5 n_{Q} + n^{2}_{Q} (6 + 125 n_{Q} )) \right]$ & 8 & \hyperref[sec:modelI]{I} ,  \hyperref[sec:modelV]{V}    \\ 
  \hline
 $u Q Q Q Q Q L e$ & $\frac{1}{24} n^{2}_{L} n^{3}_{Q} \left[-2 - n_{Q} + n^{2}_{Q} (2 + 25 n_{Q})\right]$ & 6 & -  \\ 
 \hline
 \multirow{2}{6em}{$ Q Q Q Q Q Q L L$}
  & \multirow{2}{37em}{ $\frac{1}{72} n_{L} n^{2}_{Q} \left[ 2 + 75 n_{Q} + 2 n^{2}_{Q} - n^{3}_{Q} (3 + 4 n_{Q}) + n_{L} (16 - 12 n_{Q} + 7 n^{2}_{Q} + n^{3}_{Q} (12 + 49 n_{Q})) \right]$ }& \multirow{2}{1em}{11} &   \hyperref[sec:modelI]{I} ,  \hyperref[sec:modelIII]{III}  \\ &&&    \hyperref[sec:modelV]{V} , \hyperref[sec:modelIX]{IX}  \\  [1ex] 
 \hline
\end{tabular}
\caption{Enumeration of dimension-12 operators with $\Delta B = \pm \Delta L = - 2$. The first column indicates the field content of the operator, while the second column gives the output of the Hilbert series; the number of operators for a given field content. Here, $n_{Q}$ and $n_{L}$ are the number of quark and lepton flavors, respectively. The third column gives the minimal number of terms that is required to write down an operator basis for $n_{Q}=n_{L}=3$ (which can be calculated using group-theoretic methods~\cite{Fonseca:2017lem,Fonseca:2019yya}), and the last column lists the models we consider which produce the given operator.
By operator we mean gauge and Lorentz invariant contractions of the fields with the flavor indices expanded, while a term is a collection of operators with the flavor indices unexpanded.
Also, to simplify the notation we have dropped the subscripts on the fermions.}
\label{tab:operators}
\end{center}
\end{table}



\section*{Minimal Renormalizable Models}

We now discuss the simplest models that produce some of the dimension-12 operators listed above, but don't contribute to $\Delta B = - 1$ processes and $\Delta B =-2, ~ \Delta L = 0$ processes for generic values of the couplings. 
To be more specific, we consider models with additional scalars. 
We haven't considered models with additional fermions or gauge bosons, however, they are  certainly more complicated.
By simplest we mean the lowest number of new representations. 
We find  that there are two minimal models, each containing one color sextet scalar and one color singlet scalar with non-zero hypercharge. 
The masses of these colored scalars cannot be too small in order to be consistent with 
constraints from the LHC.
In ref.~\cite{Richardson:2011df}, four-jet events from gluon fusion were
simulated. These constraints are independent of the coupling to quarks.
With no observed deviation from the standard model result, the
masses of the new colored sextet scalars have to be
\begin{equation}
        M_{X_{1}} \gtrsim 1 \quad {\rm TeV}.
        \label{eq:boundM1}
\end{equation}

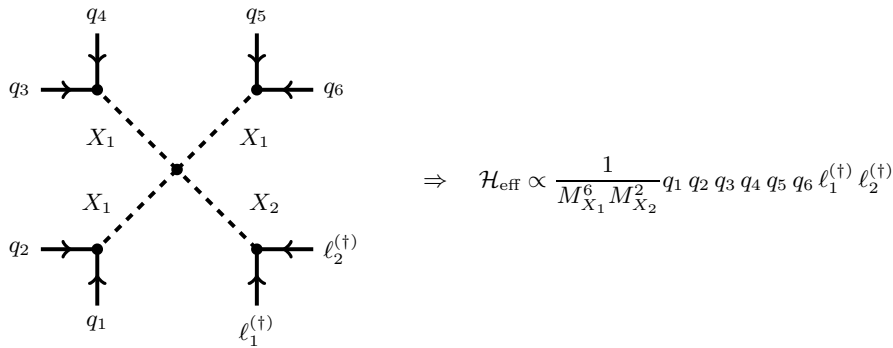
\begin{figure}
\begin{equation*}
\begin{gathered}
\begin{tikzpicture}[line width=1.5 pt,node distance=1 cm and 1.5 cm]
\coordinate(vcentral);
\coordinate[above left =  1.5cm of vcentral](v1);
\coordinate[above right =  1.5cm of vcentral](v2);
\coordinate[below right = 1.5cm of vcentral](v3);
\coordinate[below left = 1.5cm  of vcentral](v4);
\coordinate[below =  0.9 cm of v1,label=$\ X_1$](v1aux);
\coordinate[below =  0.9 cm of v2,label=$\! X_1$](v2aux);
\coordinate[above =  0.35 cm of v3,label=$\ \ X_2$](v3aux);
\coordinate[above =  0.35 cm of v4,label=$X_1$](v4aux);
\coordinate[left= 0.75 cm of v1,label=left:$q_3$](f1a);
\coordinate[above = 0.75cm of v1,label=above:$q_4$](f1b);
\coordinate[above = 0.75cm of v2,label=above:$q_5$](f2a);
\coordinate[right = 0.75 cm of v2,label=right: $q_6$](f2b);
\coordinate[left = 0.75cm of v4,label=left: $q_2$](f4a);
\coordinate[below = 0.75cm of v4,  label=below:$q_1$](f4b);
\coordinate[right= 0.75cm of v3,label=right: $\ell_2^{(\dagger)}$](f3a);
\coordinate[below = 0.75 cm of v3,label=below: $\ell_1^{(\dagger)}$](f3b);
\draw[fermion](f1a)--(v1);
\draw[fermion](f1b)--(v1);
\draw[fermion](f2a)--(v2);
\draw[fermion](f2b)--(v2);
\draw[fermion](f3a)--(v3);
\draw[fermion](f3b)--(v3);
\draw[fermion](f4a)--(v4);
\draw[fermion](f4b)--(v4);
\draw[scalarnoarrow] (vcentral) -- (v1);
\draw[scalarnoarrow] (vcentral) -- (v2);
\draw[scalarnoarrow] (v3) -- (vcentral);
\draw[scalarnoarrow] (vcentral) -- (v4);
\draw[fill=black] (vcentral) circle (.05cm);
\draw[fill=black] (v1) circle (.05cm);
\draw[fill=black] (v2) circle (.05cm);
\draw[fill=black] (v3) circle (.05cm);
\draw[fill=black] (v4) circle (.05cm);
\end{tikzpicture}
\end{gathered} \quad \quad \Rightarrow \quad {\cal H}_\text{eff} \propto \frac{1}{M_{X_{1}}^6 M_{X_{2}}^2} q_1 \, q_2 \, q_3 \, q_4 \, q_5 \, q_6 \, \ell_1^{(\dagger)}\,  \ell_2^{(\dagger)}  
\end{equation*}
\caption{\label{fig:dim12interaction}Skeleton of the renormalizable interaction that leads to a dimension-12 operator with $\Delta B = \pm \Delta L = - 2$ after integrating out the scalar mediators. Here, $q_i$ ($\ell_i$) symbolizes a quark (lepton), $X_1$ is an $SU(3)$ sextet scalar or triplet scalar, and $X_2$ is a color singlet scalar with hypercharge $\pm 1$ or $\pm 2$.}
\end{figure}

In all $\Delta B =\pm\Delta L = - 2$ operators listed above, six quarks and two leptons are involved. 
Fig.~\ref{fig:dim12interaction} illustrates the skeleton of the core renormalizable interaction that leads to the $\Delta B = \pm \Delta L = -2 $ dimension-12 operator once the scalar mediators are integrated out.

In the remainder of this section, we discuss the phenomenology of the simplest models.

\subsection*{Model I\label{sec:modelI}}

With scalars in the representations $X_{1} \sim (\bar{\bm{6}}, \bm{1})_{-1/3}$ and
$X_{2} \sim (\bm{1}, \bm{1})_{1}$, we have that
\begin{align}
	\Lagr_{\rm I} =& - g^{pr}_{1} (Q^{p}_{L\alpha} \epsilon Q^{r}_{L\beta}) X^{\alpha \beta}_{1} 
	- g^{pr}_{2} (u^{p}_{R\alpha} d^{r}_{R\beta}) X^{\alpha \beta}_{1} 
	- g^{pr}_{3} (L^{p}_{L} \epsilon L^{r}_{L}) X_{2}
	+ \lambda X^{\alpha \alpha^{\prime}}_{1} X^{\beta \beta^{\prime}}_{1} X^{\gamma \gamma^{\prime}}_{1}
	X_{2} \epsilon_{\alpha \beta \gamma} \epsilon_{\alpha^{\prime} \beta^{\prime} \gamma^{\prime}} 
	+ \textrm{h.c.}
\end{align}
where Greek letters are color indices, the superscripts $p,r, \ldots$ on the fields are flavor quantum numbers, and the quantities in round brackets are Lorentz singlets and weak $SU(2)$ singlets. Lorentz and fundamental weak $SU(2)$ indices are not displayed explicitly. 
The couplings $g_{1}$ and $g_{3}$ are antisymmetric in the flavor indices,
while $g_{2}$ has no symmetry.
We have not displayed other new scalar interactions that conserve baryon and lepton number.
This model is identical to Model $5$ in ref.~\cite{Arnold:2012sd}.

Constraints on the coupling matrices $g_1$, $g_2$, and $g_3$ arise from limits on  low-energy  flavor changing processes, LEP data, and LHC data.  For low-energy processes that occur well below the masses of the new scalars they can be integrated out giving rise to local four-fermion operators. It is convenient, for comparison with existing analysis of data, to express those  operators in terms of four-component fields. Using that notation, $X_2$ exchange gives
\begin{equation}
\label{fourLepton}
     H_{\rm eff}^{(2)} =  2\left(\frac{g_3^{pr}g_3^{* p'r'} }{ M_{X_2}^2 }\right) \left( {\bar \nu_L^{p'}} \gamma^{\mu} e_L^{p}\right) \left( {\bar e_L^{r'}} \gamma_{\mu} { \nu_L^{r}}  \right),
\end{equation}
where the round bracketed fermion bilinears are Lorentz four-vectors.  
Similarly, $X_1$ exchange gives
\begin{align}
H_{\rm eff}^{(1)} &= - \left({g_1^{pr}g_1^{* p'r'} \over M_{X_1}^2 }\right) \left[ 
 \left({\bar u_{L\alpha}^{p'}} \gamma^{\mu} {u_{L\alpha}^{p}}\right)
\left({\bar d_{L\beta}^{r'}} \gamma_{\mu} {d_{L\beta}^{r}}\right)
+ \left(
{\bar u_{L\alpha}^{p'}} \gamma^{\mu} {u_{L\beta}^{p}} \right)\left(
{\bar d_{L\beta}^{r'}} \gamma_{\mu} {d_{L\alpha}^{r}}\right)  \right]  \\
&-\left({g_2^{pr}g_2^{* p'r'} \over 4 M_{X_1}^2 }\right) \left[ \left(
{\bar u_{R\alpha}^{p'}} \gamma^{\mu} {u_{R\alpha}^{p}} \right)\left(
{\bar d_{R\beta}^{r'}} \gamma_{\mu} {d_{R\beta}^{r}}\right)  + \left(
{\bar u_{R\alpha}^{p'}} \gamma^{\mu} {u_{R\beta}^{p}} \right)\left(
{\bar d_{R\beta}^{r'}} \gamma_{\mu} {d_{R\alpha}^{r}}\right)  \right] \nonumber \\
&+\left({g_2^{pr}g_1^{* p'r'} \over  M_{X_1}^2 }\right) \left[ \left(
{\bar u_{L\alpha}^{p'}} {u_{R\alpha}^{p}} \right)\left(
{\bar d_{L\beta}^{r'}}  {d_{R\beta}^{r}}\right) + \left(
{\bar u_{L\alpha}^{p'}} {u_{R\beta}^{p}} \right)\left(
{\bar d_{L\beta}^{r'}}  {d_{R\alpha}^{r}}\right)  -\left(
{\bar d_{L\alpha}^{p'}} {u_{R\alpha}^{p}} \right)\left(
{\bar u_{L\beta}^{r'}}  {d_{R\beta}^{r}}\right)  - \left(
{\bar d_{L\alpha}^{p'}} {u_{R\beta}^{p}} \right)\left(
{\bar u_{L\beta}^{r'}}  {d_{R\alpha}^{r}}\right)  \right]
\nonumber \\ &
+\left({g_1^{pr}g_2^{* p'r'} \over  M_{X_1}^2 }\right) \left[ \left(
{\bar u_{R\alpha}^{p'}} {u_{L\alpha}^{p}} \right)\left(
{\bar d_{R\beta}^{r'}}  {d_{L\beta}^{r}}\right) + \left(
{\bar u_{R\alpha}^{p'}} {u_{L\beta}^{p}} \right)\left(
{\bar d_{R\beta}^{r'}}  {d_{L\alpha}^{r}}\right)  - \left(
{\bar u_{R\alpha}^{p'}} {d_{L\alpha}^{p}} \right)\left(
{\bar d_{R\beta}^{r'}}  {u_{L\beta}^{r}}\right)  - \left(
{\bar u_{R\alpha}^{p'}} {d_{L\beta}^{p}} \right)\left(
{\bar d_{R\beta}^{r'}}  {u_{L\alpha}^{r}}\right)  \right]. \nonumber
\end{align}
Although no tree-level flavor changing neutral currents are produced by this Hamiltonian, meson-antimeson mixing is present at one loop. For convenience, we focus on the $g_{1}$ coupling matrix between the left-handed quarks and the $X_{1}$ scalar, and we try various choices for the coupling to get an impression of the size of these constraints. We let only one of the three couplings be non-zero. 
The constraints from meson-antimeson mixing are \cite{Giudice:2011ak}
\begin{align}
    M_{X_{1}} &\gtrsim 300 |\sqrt{2} g_{1}^{pr}|^{2} ~ {\rm TeV} \qquad {\rm if }~~ g^{pr}_{1} = g^{13}_{1} \delta^{p1}\delta^{r3} ~~ {\rm or} ~~ g^{pr}_{1} = g^{23}_{1} \delta^{p2}\delta^{r3} , 
     \\
    M_{X_{1}} &\gtrsim 100 |\sqrt{2} g_{1}^{pr}| ~~ {\rm TeV} \qquad ~ {\rm if }~~ g^{pr}_{1} = g^{12}_{1} \delta^{p1}\delta^{r2} . 
\end{align}
Rotation to the mass eigenstate basis will generate non-zero entries in the coupling constant matrix, but this is a small effect.
For a sizable coupling constant $g_{1}$, this would put so strong constraints on the mass $M_{X_{1}}$ that baryon number violating processes certainly would not be observable in the laboratory.
We proceed by assuming that the $g_{1}$ coupling is very small and can be neglected.

The coupling $g_{2}$ between the $X_{1}$ scalar and the right-handed quarks can also lead to meson-antimeson mixing at one loop. We avoid these experimental constraints by assuming that the right-handed quarks are mass eigenstate fields and that $g^{pr}_{2} = g^{11}_{2} \delta^{p1} \delta^{r1}$. This assumption also avoids other flavor constraints. A nontrivial Cabibbo-Kobayashi-Maskawa (CKM)  matrix is still allowed since it arises from transforming the left-handed quark  fields to diagonalize the quark  mass matrices. 
Although we have not found any analysis of high-energy collider data constraining the color sextet scalar coupled to $u_{R} d_{R}$, other studies in the literature constraining similar diquarks could be employed to estimate the constraints in this scenario.  For instance, ref.~\cite{Pascual-Dias:2020hxo} uses LHC dijet data to constrain a diquark $D \sim (\bm{ \bar{3}}, \bm{1})_{-2/3}$ that interacts with the down-type quarks through the following interaction Lagrangian,
\begin{equation}
    {\cal L}_{{\bf \bar 3}} = - g_{D} \, D^\dagger_\alpha \, (d_{R\beta}  \, s_{R\gamma}) \epsilon^{\alpha \beta \gamma} + \text{h.c.} 
\end{equation}
leading to the following constraint,
\begin{equation}
    |g_D| \lesssim 0.1 \quad \text{ for } M_D = 1 \text{ TeV} ,
    \label{eq:boundLHC}
\end{equation}
where $M_D$ is the mass of the diquark. We use the bound in eq.~\eqref{eq:boundLHC} for $|g^{11}_{2}|$. 
 
 Next, we focus on the lepton sector. 
 There are very strong constraints on charged lepton flavor violating processes, e.g., $\mu \to e \gamma$. To avoid the stringent experimental constraints in such processes we will assume $g_3^{pr} = g^{13}_{3} \delta^{p1}\delta^{r3}$. Using the results from ref.~\cite{Crivellin:2020klg}, where in our case $\text{Br}(X_2 \to e^+ \bar \nu_\tau)\sim 50\%$, we find the following lower bound on the mass of the singly charged scalar
 \begin{equation}
     M_{X_2} \gtrsim 330 \text{ GeV}.
 \end{equation}
 Strong constraints come from the tests of lepton flavor universality. 
The coupling $g_3^{13}$ enters in the squared amplitude for the leptonic decay $\tau \rightarrow e {\bar \nu_e} { \nu_{\tau}}$ as follows
\begin{equation}
\label{eq:tautoe}
  |{\cal M}(\tau \rightarrow e {\bar \nu_e} { \nu_{\tau}})|^2=|{\cal M} (\tau \rightarrow e {\bar \nu_e} { \nu_{\tau}})_{\rm SM} |^2 \times (1+\delta_{e\tau})^2 ,
  \end{equation}
with
\begin{equation}
\delta_{e\tau} = \frac{1}{\sqrt{2}G_F} \left( \frac{|g_3^{13}|}{M_{X_2}}\right)^2 ,
\end{equation}
where in eq.~\eqref{eq:tautoe} SM means standard model. 
The bounds on lepton flavor universality sensitive to this specific element of $g_3$ are~\cite{Amhis:2019ckw}
\begin{align}
\label{tautomu}
\left(\frac{\text{Br}( \tau \to \mu \bar \nu_\mu \nu_\tau)/ \text{Br}(\tau \to \mu \bar \nu_\mu \nu_\tau)_\text{SM}}{\text{Br}(\tau \to e \bar \nu_e \nu_\tau)/ \, \text{Br}(\tau \to e \bar \nu_e \nu_\tau)_\text{SM}}\right)^{\frac{1}{2}} &= (1+\delta_{e\tau})^{-1} \simeq 1 - \delta_{e\tau} \leq 1.0018(14) ,  \\
\label{tautoe}
 \left(\frac{\text{Br}(\tau \to e \bar \nu_e \nu_\tau)/\text{Br}(\tau \to e \bar \nu_e \nu_\tau)_\text{SM}}{\text{Br}(\mu \to e \bar \nu_e \nu_\mu)/\text{Br}(\mu \to e \bar \nu_e \nu_\mu)_\text{SM}}\right)^{\frac{1}{2}} &= \ (1+\delta_{e\tau}) \simeq 1 + \delta_{e\tau} \leq 1.0010(14) .
\end{align}
The strongest constraint is set by eq.~\eqref{tautomu}. At $2\sigma$, 
\begin{equation}
    |g_3^{13}| \lesssim 0.13 \left(\frac{M_{X_2}}{\text{TeV}}\right) .
\end{equation}
The $g_{3}^{13}$ coupling also enters in the expression for the lepton magnetic moments, leading to a weaker constraint than the tests on lepton flavor universality do.

From this and the other experimental constraints discussed above, we will choose some allowed values of the masses and coupling constants to estimate the size of baryon number violating processes.

The dominant subprocess violating $\Delta B = \Delta L = -2$ in this model is the dinucleon decay $n \, p \to e^+ \, \bar\nu$.
The effective Hamiltonian that leads to this dinucleon decay is
\begin{equation}
H_{\rm eff}^{\Delta B=\Delta L=-2}= - \left({ \lambda^{*} \over M_{X_1}^{6} M_{X_2}^{2}}\right)g_{2}^{pr}g_{2}^{p'r'}g_{2}^{p''r''}g_3^{st}(u_{R\{\alpha}^pd_{R\alpha'\}}^r)(u_{R\{\beta}^{p'}d_{R\beta'\}}^{r'})(u_{R\{\gamma}^{p''}d_{R\gamma'\}}^{r''})(L_{L}^{ s} \epsilon L_L^{t})\epsilon^{\alpha \beta \gamma}\epsilon^{\alpha' \beta' \gamma'} ,
\label{eq:interactionB2L2MI}
\end{equation}
where the brace brackets mean symmetrization of the color indices.
The rate for the dinucleon decay in a nucleus can be estimated from the cross section $\sigma(n  p \to e^+  \bar\nu)$ as follows~\cite{Goity:1994dq}:
\begin{equation}
\label{eq:decayrate}
    \Gamma_{np} = \frac{1}{(2\pi)^3\rho_N}\int d^3 k_1 \, d^3 k_2 \, \rho_N(k_1) \, \rho_N(k_2) \, v_\text{rel} \, (1-\vec{v}_{1}\cdot \vec{v}_{2})\, \sigma(n \, p \to e^{+} \bar\nu) ,
\end{equation}
where $\rho_N = \int d^3 k \, \rho_N(k)/\sqrt{(2\pi)^3}$ is the average nuclear density, $\rho_N \sim 0.25 \text{ fm}^{-3}$, and $\vec{v}_{1}$ and $\vec{v}_{2}$ are the nucleon velocities, which we presume small.
The lifetime of the dinucleon decay ($1/\Gamma_{np}$) is therefore estimated to be
\begin{equation}
\begin{split}
    \tau_{n p \to e^+ \bar\nu} & \sim 32 \pi \frac{m_N^2}{\rho_N} |g_{2}^{11}|^{-6}|g_{3}^{13}|^{-2}|\lambda|^{-2}\frac{M_{X_1}^{12}M_{X_2}^{4}}{\Lambda_\text{QCD}^{16}}\\
    & \sim 1.65 \times 10^{42}\text{ years} \left(\frac{0.1}{|g_2^{11}|}\right)^6 \left(\frac{0.01}{|g_{3}^{13}|}\right)^{2}\left(\frac{1}{|\lambda|}\right)^2 \left( \frac{M_{X_1}^{12}M_{X_2}^4}{\text{TeV}^{16}}\right) .
    \end{split}
\end{equation}
Using the following values for the couplings and masses consistent with the experimental constraints, $|g_2^{11}| = 0.1$, $|g_{3}^{13}| = 0.04$, $|\lambda|=2$, $M_{X_1}=1$ TeV, $M_{X_2} = 350$ GeV, the estimate for the lifetime is 
\begin{equation}
    \tau_{np \to e^+ \bar \nu} \sim 3.9 \times 10^{38} \text{ years}.
\end{equation}
The bound given by the Super-Kamiokande collaboration is $\tau_{np \to e^+  \bar\nu}>2.6 \times 10^{32} \text{ years}$~\cite{Takhistov:2015fao}. Even given the large uncertainties in our estimate of the hadronic matrix element, in this model it is unlikely that baryon number violating processes will be observed in the laboratory.

\subsection*{Model II\label{sec:modelII}}

In the second minimal renormalizable model, the new scalars in the representations $X_{1} \sim (\bar{\bm{6}}, \bm{1})_{2/3}$ and
$X_{2} \sim (\bm{1}, \bm{1})_{2}$ only couple to right-handed fermions.
The new renormalizable interactions are
\begin{align}
	\Lagr_{\rm II} =& - g^{pr}_{1} (d^{p}_{R\alpha} d^{r}_{R\beta}) X^{\alpha \beta}_{1} 
	- g^{pr}_{2} (e^{p}_{R} e^{r}_{R}) X_{2}
	+ \lambda X^{\alpha \alpha^{\prime}}_{1} X^{\beta \beta^{\prime}}_{1} X^{\gamma \gamma^{\prime}}_{1}
	X^{\dagger}_{2} \epsilon_{\alpha \beta \gamma} \epsilon_{\alpha^{\prime} \beta^{\prime} \gamma^{\prime}} 
	+ \textrm{h.c.}
\end{align}
The couplings $g_{1}$ and $g_{2}$ are symmetric in the flavor indices.
This model is identical to Model $8$ in ref.~\cite{Arnold:2012sd}. 

At low energies the effective Hamiltonian for baryon and lepton number violating processes is
\begin{equation}
H_{\rm eff}^{\Delta B= -\Delta L= -2}= - \left({ \lambda^{*} \over M_{X_1}^{6} M_{X_2}^{2}}\right)g_1^{pr}g_1^{p'r'}g_1^{p''r''}g_2^{*st}(d_{R\alpha}^pd_{R\alpha'}^r)(d_{R\beta}^{p'}d_{R\beta'}^{r'})(d_{R\gamma}^{p''}d_{R\gamma'}^{r''})(e_R^{\dagger s}e_R^{\dagger t})\epsilon^{\alpha \beta \gamma}\epsilon^{\alpha' \beta' \gamma'} .
\label{eq:interactionB2L2MIII}
\end{equation}
Exchanges of the $X_1$ and $X_2$ scalars give local four-fermion interactions that conserve baryon and lepton number. In terms of four-component fields, they are
\begin{equation}
\label{fourlepton}
 H_{\rm eff}^{(1)}= - \left({g_1^{pr}g_1^{* p'r'} \over 2 M_{X_1}^2 }\right) \left( {\bar d_{R\alpha}^{p'} } \gamma^{\mu} d_{R\alpha}^p \right) \left( {\bar d_{R\beta}^{r'} } \gamma_{\mu} d_{R\beta}^r\right) , ~~  \quad   
H_{\rm eff}^{(2)}= - \left({g_2^{pr}g_2^{* p'r'} \over 2 M_{X_2}^2 }\right) \left( {\bar e_R^{p'} } \gamma^{\mu} e_R^p \right) \left( {\bar e_R^{r'} } \gamma_{\mu} e_R^r\right).
\end{equation}
Flavor changing effects are avoided by assuming that the coupling constant matrices $g_{1}$ and  $g_{2}$  have only one non-zero, diagonal entry and that the right-handed standard model fields are mass eigenstates. As noted earlier, nontrivial CKM and Pontecorvo–Maki–Nakagawa–Sakata (PMNS) matrices are still allowed since they arise from transforming the left-handed quark and lepton fields to diagonalize the quark and lepton mass matrices. 

For the colored scalar interaction with down-type quarks we adopt $g^{pr}_1 = g^{11}_1\delta^{p1}\delta^{r1}$ as the only non-zero entry of the matrix in flavor space. 
Again, although Model II contains a color sextet, and not a color anti-triplet, we use the constraint on $g_{D}$ in eq.~\eqref{eq:boundLHC} for $|g^{11}_{1}|$.

Searches for doubly charged scalars generated by pair production $pp \to \gamma^* \to X_2 X_2^*$ and photon fusion can set a lower bound on the mass of the scalar $X_2$ that couples to the leptons. In particular, the analysis by ATLAS~\cite{Aaboud:2017qph} gives the following bound on the mass of a doubly charged scalar that only couples to right-handed charged leptons:
\begin{equation}
\label{eq:massX2}
M_{X_2} \gtrsim 660 \text{ GeV},
\end{equation}
under the assumption that $\text{Br}(X_2 \to \ell^+ \ell^+) = 100\%$, where $\ell = e \text{ or } \mu$.
In addition, there is a strong bound from LEP II $e^+e^-$ annihilation data~\cite{Berthier:2015gja}, 
\begin{equation}
   |g_2^{11}|\lesssim 0.18 \left(\frac{M_{X_2}}{{\rm TeV}}\right) . 
\end{equation}
This experimental bound can be evaded by having a small $g^{11}_{2}$ coupling.
We assume $g^{pr}_{2} = g^{22}_{2} \delta^{p2} \delta^{r2}$, thus evading experimental constraints from LEP and lepton flavor violating decays. In this case, the low-energy baryon number violating processes will have final state muons (e.g., $nn \rightarrow \pi^+ \pi^+ \mu \mu$). The $g_2^{22}$ coupling contributes to the muon magnetic moment, 
\begin{equation}
    \Delta a_\mu (X_2) \simeq - \frac{m_\mu^2}{6\pi^2}\left(\frac{|g_2^{22}|}{M_{X_2}}\right)^2  .
\end{equation}
The current discrepancy between the standard model prediction and the experimental measurement of the muon anomalous magnetic moment adds up to $3.3$ times the combined theoretical and experimental error. The new interactions in this model cannot be the new physics that explains this discrepancy because the contribution from the doubly charged scalar to $a_\mu$ is negative. Therefore, we assume that the standard model prediction will eventually match the experimental measurement and impose that the shift in $a_\mu$ induced by $X_2$
lies in the 2$\sigma$ window of the experimental value 
$a_\mu^\text{exp} = 11659209.1(5.4)(3.3)\times 10^{-10}$~\cite{Bennett:2006fi,RevModPhys.84.1527}. The later imposes the following constraint on the $g_2^{22}$ and $M_{X_2}$ parameters,
\begin{equation}
    |g_2^{22}|\lesssim 2.6 \left(\frac{M_{X_2}}{\text{TeV}}\right).
\end{equation}

We combine the constraints discussed above to estimate the lifetime for the $\Delta B = - \Delta L = -2$ processes consistent with experimental constraints.
The dominant subprocess violating $\Delta B = - \Delta L = -2$ in this model is the dinucleon decay $n  n \to \pi^+  \pi^+  \mu \mu$. 
The dinucleon decay rate in a nucleus can be computed from eq.~\eqref{eq:decayrate} with the cross section $\sigma(n  n \rightarrow \pi^+  \pi^+   \mu  \mu)$. 
The lifetime of the dinucleon decay is therefore estimated to be
\begin{equation}
\begin{split}
    \tau_{nn} & \sim \frac{2^{15} \, 3 \pi^5}{\rho_N \, m_N^2} |g_1^{11}|^{-6}|g_2^{22}|^{-2}|\lambda|^{-2} \frac{M_{X_1}^{12}M_{X_2}^4}{\Lambda_\text{QCD}^{12}}\\
    &\sim 7.9 \times 10^{40}\text{ years} \times \left(\frac{0.1}{|g_1^{11}|}\right)^6 \left(\frac{1}{|g_2^{22}|}\right)^2\left(\frac{1}{|\lambda|}\right)^2\left(\frac{M_{X_1}^{12}M_{X_2}^4}{\text{TeV}^{16}}\right),
    \end{split}
\end{equation}
where a four-body massless phase space has been inserted~\cite{Asatrian:2012tp}. Taking the reasonable values for the couplings $|g_2^{22}| =1$,  $|\lambda | = 2$, and $|g_{1}^{11}| = 0.1$, and the lowest values allowed for the scalar masses, $M_{X_1} = 1 \text{ TeV}$ and $M_{X_2} = 660 \text{ GeV}$, we obtain the following estimate for the lifetime:
\begin{equation}
    \tau_{nn \to \pi^+\pi^+\mu^-\mu^-} \sim 3.7 \times 10^{39} \text{ years}.
\end{equation}
No direct experimental searches for such dinucleon decays have been performed. However, inclusive searches could capture the relevant decays. Current limits from inclusive nucleon decay searches are quite old, but could potentially be improved to $\tau = \mathcal{O}(100) \times 10^{30}$ years \cite{Heeck:2019kgr}. 
The use of four-body
phase space may underestimate the rate for baryon number violating processes in this model since the final state pions can be virtual and give rise to decays like $(A,Z)\rightarrow (A-2, Z+2) +\mu \mu.$
Despite this uncertainty, our conclusion regarding the observability of baryon number violating processes in the laboratory for this model is similar to Model I.

\section*{Non-minimal Renormalizable Models}

We now list some non-minimal renormalizable models containing scalars. 
We allow for scalars in more than two representations and require that the leading baryon number violating processes have $\Delta B = \pm \Delta L = -2$ as before.

\subsection*{Model III\label{sec:modelIII}}

In this model, we add scalars in the representations $X_{1} \sim (\bar{\bm{6}}, \bm{3})_{-1/3}$ and
$X_{2} \sim (\bm{1}, \bm{1})_{1}$. 
The new interactions are
\begin{align}
	\Lagr_{\rm III} =& - g^{pr}_{1} (Q^{p}_{L\alpha} \epsilon \tau^{A} Q^{r}_{L\beta}) X^{\alpha \beta A}_{1} 
	- g^{pr}_{2} (L^{p}_{L} \epsilon L^{r}_{L}) X_{2}
	+ \lambda X^{\alpha \alpha^{\prime}A}_{1} X^{\beta \beta^{\prime}B}_{1} X^{\gamma \gamma^{\prime}C}_{1}
	X_{2} \epsilon_{\alpha\beta\gamma} \epsilon_{\alpha^{\prime} \beta^{\prime} \gamma^{\prime}} \epsilon^{ABC}
	+ \textrm{h.c.}
\end{align}
where we have three different $X_{1}$'s which are needed for the $\lambda$ interaction
to be non-zero.
 We use capital Latin letters for weak adjoint indices.
The couplings $g_{1}$ and $g_{2}$ are symmetric and antisymmetric in the flavor indices, respectively.
This model is identical to Model $6$ in ref.~\cite{Arnold:2012sd}.

 The discussion of the lepton sector of this model is the same as in Model I.

Neglecting weak symmetry breaking in the $X_1$ masses, $X_1$ exchange gives
\begin{align}
\label{4quark}
H_{\rm eff}^{(1)}=&-\left({g_1^{pr}g_1^{* p'r'} \over 2 M_{X_1}^2 }\right) \left[ \left(
{\bar Q_{L\alpha}^{p'}} \gamma^{\mu} {Q_{L\alpha}^{p}} \right)\left(
{\bar Q_{L\beta}^{r'}} \gamma_{\mu} {Q_{L\beta}^{r}}\right) +\left(
{\bar Q_{L\alpha}^{p'}} \gamma^{\mu} {Q_{L\beta}^{p}} \right)\left(
{\bar Q_{L\beta}^{r'}} \gamma_{\mu} {Q_{L\alpha}^{r}}\right)  \right] .
\end{align}
The quark fields in eq.~\eqref{4quark} are not mass eigenstates. To transform to mass eigenstate fields, we use
\begin{equation}
u_L^p\rightarrow U^{pq}(u,L) u_L^q,~\quad ~d_L^p\rightarrow U^{pq}(d,L) d_L^q .
\end{equation}
Flavor constraints on the scalar mass $M_{X_{1}}$ and coupling $g_{1}$ are very strong. We explore several possible choices for the couplings in an attempt to minimize them.

The part of 
$H_{\rm eff}^{(1)}$ that gives rise to flavor changing neutral currents expressed in terms of the mass eigenstate fields is,
\begin{align}
\Delta H^{(1)}_{\rm eff}=&-\left({{\tilde g}_1^{pr}{\tilde g}_1^{* p'r'} \over 2 M_{X_1}^2 } \right) \left[ \left(
{\bar u_{L\alpha}^{p'}} \gamma^{\mu} {u_{L\alpha}^{p}} \right)\left(
{\bar u_{L\beta}^{r'}} \gamma_{\mu} {u_{L\beta}^{r}}\right) +\left(
{\bar u_{L\alpha}^{p'}} \gamma^{\mu} {u_{L\beta}^{p}} \right)\left(
{\bar u_{L\beta}^{r'}} \gamma_{\mu} {u_{L\alpha}^{r}}\right)  \right]  
\nonumber \\ 
&-\left({{\hat g}_1^{pr}{\hat g}_1^{* p'r'} \over 2 M_{X_1}^2 } \right) \left[ \left(
{\bar d_{L\alpha}^{p'}} \gamma^{\mu} {d_{L\alpha}^{p}} \right)\left(
{\bar d_{L\beta}^{r'}} \gamma_{\mu} {d_{L\beta}^{r}}\right) +\left(
{\bar d_{L\alpha}^{p'}} \gamma^{\mu} {d_{L\beta}^{p}} \right)\left(
{\bar d_{L\beta}^{r'}} \gamma_{\mu} {d_{L\alpha}^{r}}\right)  \right]  .
\end{align}
Here,
\begin{equation}
{\tilde g}_1^{pr} =g_1^{st}  U(u,L)^{sp}U(u,L)^{tr} , 
\qquad
 {\hat g}_1^{pr} =g_1^{st}  U(d,L)^{sp}U(d,L)^{tr} .
\end{equation}
The mass eigenstate couplings ${\tilde g}_1^{pr}$ and ${\hat g}_1^{pr}$ are also symmetric in flavor. While it may be possible to choose $g_1$ and the rotation matrices so that both ${\tilde g}_1$ and ${\hat g}_1$ are flavor diagonal with only one non-zero entry, that seems very contrived. A further constraint is that we don't want both ${\tilde g}_1^{11}$ and ${\hat g}_1^{11}$ to be suppressed by small weak mixing angles since that would suppress the rate for laboratory baryon number violating processes. 

To explore this further, let us imagine that the left-handed up-type quark fields are actually mass eigenstate fields; $U^{pq}(u,L)= \delta^{pq}$. Then $U(d,L)$ is the CKM matrix $V_{\rm CKM}$. Flavor changing neutral currents can then be avoided in the up-quark sector by assuming that the coupling constant $g_{1}$ have only one non-zero, diagonal entry. 
Suppose we take $g_1^{pq}=g^{11}_1 \delta^{p1}\delta^{q1}$. To leading order in small mixing angles the effective Hamiltonian for $K-{\bar K}$ meson mixing is:
\begin{equation}
\label{KK}
\Delta H_{K-{\bar K}}\simeq -\left({ |g_1^{11}|^2 s_{12}^2 \over 2 M_{X_1}^2} \right) \left[ \left({\bar d_{La}} \gamma^{\mu} s_{La} \right)\left({\bar d_{Lb}} \gamma_{\mu} s_{Lb} \right) + \left({\bar d_{La}} \gamma^{\mu} s_{Lb} \right)\left({\bar d_{Lb}} \gamma_{\mu} s_{La} \right) + {\rm h.c. }
 \right] ,
\end{equation}
where we are using the convention for the CKM matrix used by the particle data group~\cite{Zyla:2020zbs}. This effective Hamiltonian implies that
\begin{equation}
\label{DeltaKK}
| \Delta m_{K^0}|/m_K \sim \left({ |g_1^{11}|^2 s_{12}^2 f_K^2\over 2 M_{X_1}^2}  \right) .
\end{equation}
Experimentally, $| \Delta m_{K^0}|/m_K \sim 10^{-14}$.
Clearly the constraint from $K^0-{\bar K}^0$ mixing on ${g_1^{11}}$ and $M_{X_1}$ that follow from eq.~(\ref{DeltaKK}) is too strong for Model III to give observable baryon number violation in the laboratory.

If we take the left-handed down-type quarks to be mass eigenstates, then the CKM matrix arises completely from redefining the up-type quarks to diagonalize their mass matrix. In that case with the assumption that $g_1^{pr}=g^{11}_1 \delta^{p1}\delta^{r1}$ 
\begin{equation}
\label{DeltaDD}
| \Delta m_{D^0}|/m_D \sim \left({ |g_1^{11}|^2 s_{12}^2 f_D^2\over 2 M_{X_1}^2}  \right) .
\end{equation}
Experimentally, $| \Delta m_{D^{0}}|/m_D \sim 10^{-15}$. 
Now constraints from $D^0-{\bar D}^0$ mixing on $g_1^{11}$ and $M_{X_1}$ prevent Model III from giving rise to observable baryon number violation in the laboratory. 

In summary, our conclusion for Model III is that, without some very contrived flavor dependence of the coupling matrix $g_1^{pq}$, limits on flavor changing neutral current processes rule out this model giving rise to observable baryon number violation in the laboratory.

\subsection*{Model IV\label{sec:modelIV}}

Next, we consider $X_{1} \sim (\bar{\bm{6}}, \bm{3})_{-1/3}$, 
$X_{2} \sim (\bar{\bm{6}}, \bm{1})_{-4/3}$, and
$X_{3} \sim (\bm{1}, \bm{1})_{2}$:
\begin{align}
	\Lagr_{\rm IV} =& - g^{pr}_{1} (Q^{p}_{L\alpha} \epsilon \tau^{A} Q^{r}_{L\beta}) X^{\alpha \beta A}_{1} 
	- g^{pr}_{2} (u^{p}_{R\alpha} u^{r}_{R\beta}) X^{\alpha \beta}_{2}
	- g^{pr}_{3} (e^{p}_{R} e^{r}_{R}) X_{3}
	+ \lambda X^{\alpha \alpha^{\prime}A}_{1} X^{\beta \beta^{\prime}A}_{1} X^{\gamma \gamma^{\prime}}_{2}
	X_{3} \epsilon_{\alpha \beta \gamma} \epsilon_{\alpha^{\prime} \beta^{\prime} \gamma^{\prime}} 
	+ \textrm{h.c.}
\end{align}
The couplings $g_{1}$, $g_{2}$, and $g_{3}$ are symmetric in the flavor indices. 
Similar to Model III, this model is constrained by meson-antimeson mixing, and thus is excluded from producing observable 
baryon number violating processes in the laboratory.

\subsection*{Model V\label{sec:modelV}}

Alternatively, $X_{1} \sim (\bar{\bm{6}}, \bm{3})_{-1/3}$, 
$X_{2} \sim (\bar{\bm{6}}, \bm{1})_{-1/3}$, and
$X_{3} \sim (\bm{1}, \bm{1})_{1}$. 
\begin{align}
	\Lagr_{\rm V} =& - g^{pr}_{1} (Q^{p}_{L\alpha} \epsilon \tau^{A} Q^{r}_{L\beta}) X^{\alpha \beta A}_{1} 
	- g^{pr}_{2} (Q^{p}_{L\alpha} \epsilon Q^{r}_{L\beta}) X^{\alpha \beta}_{2} 
	- g^{pr}_{3} (u^{p}_{R\alpha} d^{r}_{R\beta}) X^{\alpha \beta}_{2}
	- g^{pr}_{4} (L^{p}_{L} \epsilon L^{r}_{L}) X_{3} \nn
		      &+ \lambda_{1} X^{\alpha \alpha^{\prime}A}_{1} X^{\beta \beta^{\prime}B}_{1} X^{\gamma\gamma^{\prime}C}_{1} X_{3} \epsilon_{\alpha\beta\gamma} \epsilon_{\alpha^{\prime} \beta^{\prime} \gamma^{\prime}} \epsilon^{ABD}	
		    + \lambda_{2} X^{\alpha\alpha^{\prime}}_{2} X^{\beta \beta^{\prime}}_{2} X^{\gamma\gamma^{\prime}}_{2} X_{3} \epsilon_{\alpha \beta \gamma} \epsilon_{\alpha^{\prime} \beta^{\prime} \gamma^{\prime}} \nn
		      &+ \lambda_{3} X^{\alpha \alpha^{\prime}A}_{1} X^{\beta \beta^{\prime}A}_{1} X^{\gamma \gamma^{\prime}}_{2} X_{3} \epsilon_{\alpha \beta \gamma} \epsilon_{\alpha^{\prime} \beta^{\prime} \gamma^{\prime}} 
	+ \textrm{h.c.}
\end{align}
The coupling $g_{1}$ is symmetric and $g_{2}$ and $g_{4}$ are antisymmetric in the flavor indices, while
$g_{3}$ doesn't have any symmetry.
Note that this model reduces to Model III when we remove $X_{2}$,
and it reduces to Model I when we remove $X_{1}$.
The same experimental constraints apply as in Models I and III.

\subsection*{Model VI\label{sec:modelVI}}

The next model we consider is $X_{1} \sim (\bar{\bm{6}}, \bm{1})_{-4/3}$, 
$X_{2} \sim (\bar{\bm{6}}, \bm{1})_{-1/3}$, and
$X_{3} \sim (\bm{1}, \bm{1})_{2}$. 
\begin{align}
	\Lagr_{\rm VI} =& - g^{pr}_{1} (u^{p}_{R\alpha} u^{r}_{R\beta}) X^{\alpha\beta}_{1} 
	- g^{pr}_{2} (Q^{p}_{L\alpha} \epsilon Q^{r}_{L\beta}) X^{\alpha\beta}_{2}
	- g^{pr}_{3} (u^{p}_{R\alpha} d^{r}_{R\beta}) X^{\alpha\beta}_{2}
	- g^{pr}_{4} (e^{p}_{R} e^{r}_{R}) X_{3} \nn &
	+ \lambda X^{\alpha\alpha^{\prime}}_{1} X^{\beta\beta^{\prime}}_{2} X^{\gamma\gamma^{\prime}}_{2} X_{3} \epsilon_{\alpha\beta\gamma} \epsilon_{\alpha^{\prime} \beta^{\prime} \gamma^{\prime}} 
	+ \textrm{h.c.}
\end{align}
The couplings $g_{1}$ and $g_{4}$ are symmetric in the flavor indices, $g_{2}$ 
is antisymmetric,
while $g_{3}$ has no symmetry.

The effective Hamiltonian for baryon and lepton number violating processes is
\begin{equation}
H_{\rm eff}^{\Delta B= \Delta L= -2}= - \left({ \lambda^{*} \over M_{X_1}^{2}M_{X_2}^4 M_{X_3}^{2}}\right)g_1^{pr}g_3^{p'r'}g_3^{p''r''}g_4^{st}(u_{R\alpha}^pu_{R\alpha'}^r)(u_{R\{\beta}^{p'}d_{R\beta'\}}^{r'})(u_{R\{\gamma}^{p''}d_{R\gamma'\}}^{r''})(e_R^{ s}e_R^{ t})\epsilon^{\alpha \beta \gamma}\epsilon^{\alpha' \beta' \gamma'} + \dots
\label{eq:interactionB2L2MVI}
\end{equation}
We take the right-handed fermions be mass eigenstate fields and, as discussed in the other models, make $g^{pr}_{1} = g^{11}_{1} \delta^{p1} \delta^{r1}$, $g^{pr}_{3} = g^{11}_{3} \delta^{p1} \delta^{r1}$, and $g^{pr}_{4} = g^{22}_{4} \delta^{p2} \delta^{r2}$, with $g_{2}$ being small to satisfy LHC, LEP, and flavor constraints.
The lifetime for the dominant subprocess $pp \rightarrow \mu^{+} \mu^{+}$ is
\begin{equation}
\begin{split}
    \tau_{p p \to \mu^+ \mu^+} & \sim 32 \pi \frac{m_N^2}{\rho_N} |g_{1}^{11}|^{-2}|g_{3}^{11}|^{-4}|g_{4}^{22}|^{-2}|\lambda|^{-2}\frac{M_{X_1}^{4}M_{X_2}^{8}M_{X_3}^4}{\Lambda_\text{QCD}^{16}}\\
    & \sim 1.65 \times 10^{38}\text{ years} \left(\frac{0.1}{|g_1^{11}|}\right)^2 \left(\frac{0.1}{|g_3^{11}|}\right)^4 \left(\frac{1}{|g_{4}^{22}|}\right)^{2}\left(\frac{1}{|\lambda|}\right)^2 \left( \frac{M_{X_1}^{4}M_{X_2}^8M_{X_3}^4}{\text{TeV}^{16}}\right) .
    \end{split}
\end{equation}
Using the following values for the couplings and masses, 
$|g_1^{11}|=|g_3^{11}|=0.1$, $|g_4^{22}|=1$, $|\lambda|=2$, $M_{X_1}= M_{X_2}=1$ TeV, and $M_{X_3} = 660$  GeV, the lifetime is estimated to be 
\begin{equation}
\tau_{pp \to \mu^+ \mu^+} \sim 7.8 \times 10^{36} \text{ years}.
\end{equation} 
The limit from the Super-Kamiokande collaboration is $\tau_{pp \to \mu^+ \mu^+} > 4.4 \times 10^{33} \text{ years}$~\cite{Sussman:2018ylo}.
If our estimate of the hadronic matrix element using naive dimensional analysis underestimates its size by an order of magnitude, this processes might eventually be observable in the laboratory.

\subsection*{Model VII\label{sec:modelVII}}

In this model we add scalars in the representations, $X_{1} \sim (\bm{3}, \bm{1})_{2/3}$ and
$X_{2} \sim (\bm{1}, \bm{1})_{2}$. 
\begin{align}
    \Lagr_{\rm VII} &= - g^{pr}_{1}  (d^{p}_{R\alpha} d^{r}_{R\beta}) X_{1 \gamma} \epsilon^{\alpha\beta\gamma}
    - g^{pr}_{2} (e^{p}_{R} e^{r}_{R}) X_{2}
    + \lambda X_{1\alpha} X_{1\beta} X_{1\gamma} X^{\dagger}_{2} \epsilon^{\alpha\beta\gamma}
    + {\rm h.c.}
\end{align}
The coupling $g_{1}$ is antisymmetric in the flavor indices, while the coupling $g_{2}$ is symmetric.
Three different $X_{1}$ scalars are required to have a nonzero $\lambda$ interaction (due to the antisymmetric color structure). This is why the simplest models don't include this model; it contains four additional scalar representations while the simplest models only have two new scalar representations. This model is identical to Model 9 in ref.~\cite{Arnold:2012sd}.
The constraints on the couplings and masses in this model are similar to the constraints in Model II.

\subsection*{Model VIII\label{sec:modelVIII}}

Lastly, $X_{1} \sim (\bm{3}, \bm{1})_{2/3}$, $X_{2} \sim (\bar{\bm{6}}, \bm{1})_{2/3}$, and
$X_{3} \sim (\bm{1}, \bm{1})_{2}$. 
\begin{align}
    \Lagr_{\rm VIII} &= - g^{pr}_{1}  (d^{p}_{R\alpha} d^{r}_{R\beta}) X_{1 \gamma} \epsilon^{\alpha\beta\gamma}
    - g^{pr}_{2}  (d^{p}_{R\alpha} d^{r}_{R\beta}) X^{\alpha\beta}_{2}
    - g^{pr}_{3} (e^{p}_{R} e^{r}_{R}) X_{3}
    \nonumber \\ &
    + \lambda_{1} X_{1\alpha} X_{1\beta} X_{1\gamma} X^{\dagger}_{3} \epsilon^{\alpha\beta\gamma}
    + \lambda_{2} X^{\alpha\alpha^{\prime}}_{2} X^{\beta \beta^{\prime}}_{2} X^{\gamma\gamma^{\prime}}_{2} X^{\dagger}_{3} \epsilon_{\alpha\beta\gamma} \epsilon_{\alpha^{\prime}\beta^{\prime}\gamma^{\prime}}
    + \lambda_{3} X_{1\alpha} X_{1\beta} X^{\alpha\beta}_{2} X^{\dagger}_{3} 
    + {\rm h.c.}
\end{align}
The couplings $g_{2}$ and $g_{3}$ are symmetric in the flavor indices, and $g_{1}$ is antisymmetric. Models II and VII are subsets of this model.

\section*{Another Renormalizable Model}

\subsection*{Model IX\label{sec:modelIX}}

We add scalars in two different representations: $X_{1} \sim (\bar{\bm{6}}, \bm{3})_{-1/3}$ and
$X_{2} \sim (\bm{1}, \bm{3})_{1}$. 
The new interactions are
\begin{align}
    \label{eq:modelIX}
	\Lagr_{\rm IX} =& - g^{pr}_{1} (Q^{p}_{L\alpha} \epsilon \tau^{A} Q^{r}_{L\beta}) X^{\alpha \beta A}_{1} 
	- g^{pr}_{2} (L^{p}_{L} \epsilon \tau^{A} L^{r}_{L}) X^{A}_{2}
	+ \lambda X^{\alpha \alpha^{\prime}A}_{1} X^{\beta \beta^{\prime}B}_{1} X^{\gamma \gamma^{\prime}C}_{1}
	X^{D}_{2} \epsilon_{\alpha\beta\gamma} \epsilon_{\alpha^{\prime} \beta^{\prime} \gamma^{\prime}} \delta^{(AB}\delta^{C)D}
	+ \textrm{h.c.}
\end{align}
where $\delta^{(AB}\delta^{C)D} = \delta^{AB}\delta^{CD} + \delta^{BC}\delta^{AD} + \delta^{CA} \delta^{BD}$. 
The couplings $g_{1}$ and $g_{2}$ are symmetric in the flavor indices. 
We have not displayed other new scalar interactions that conserve baryon and lepton number.
This model is identical to Model $7$ in ref.~\cite{Arnold:2012sd}.

This model does not fulfil our criteria where the leading baryon number violating processes have $\Delta B = \pm \Delta L = - 2$. The scalar $X_{2}$ will have a vacuum expectation value for general values of the scalar mass parameter $\mu$, coming from the trilinear term in the scalar potential
\begin{align}
   \mu (H \epsilon \tau^{A} H) X^{A*}_{2}  + {\rm h.c.}
\end{align}
This would lead to $\Delta B = -2, ~ \Delta L = 0$ processes, which we don't consider here.
Although this model does not satisfy our criteria, nevertheless, this model could still have interesting phenomenology if $\mu$ is small enough.

\section*{Concluding Remarks}

There is no evidence of baryon number violation from laboratory experiments despite heroic efforts to observe it. If one classifies the non-renormalizable operators composed of standard model fields that can give rise to such processes in terms of the change in baryon number $\Delta B$, then it is only operators with $|\Delta B| \le 2$ that have a hope of being observed in the laboratory and not be in conflict with data. The reason for this is that models with $|\Delta B | \geq 3$ must have new degrees of freedom with masses below the weak scale for such processes to be observable in the laboratory. 

In this paper we constructed the simplest models that can give rise to $\Delta B=\pm \Delta L=-2$ but do not, for generic values of the couplings, give rise to  $\Delta B=-2, ~\Delta L=0$ or $\Delta B=  - 1$ processes. Models with $\Delta B=-2,~ \Delta L=0$ have previously been studied. We also discussed some non-minimal models and enumerated the dimension-12 operators that can give rise to $\Delta B=\pm \Delta L=-2$ processes using Hilbert series techniques.

We found that the simplest models are strongly constrained by LHC, LEP, and flavor physics. The model which gives an estimated rate of dinucleon decay closest to the experimental bound is Model VI. This model is non-minimal because it contains scalars in three representations.

 In the models we presented in this paper, lifetimes for the $\Delta B=\pm \Delta L=-2$ processes are proportional to  the twelfth power of the colored scalar masses. Improvements in our understanding of the compatibility  of these models with observable laboratory baryon number violation can be made if LHC constraints are improved. Specifically, analyses of three simplified models with colored sextets can made with the full available data set. One has a new colored sextet scalar $X^{\alpha \beta}$ with interaction Lagrange density,
\begin{equation}
    {\cal L}_{\rm int}= -g(d_{R \alpha} d_{R \beta})X^{\alpha \beta} + \text{h.c.}
\end{equation}
where the charged  scalar $X$ only couples to the down quark (i.e., not the strange or the bottom),
\begin{equation}
    {\cal L}_{\rm int}= -g(u_{R \alpha} u_{R \beta})X^{\alpha \beta} + \text{h.c.}
\end{equation}
where the charged  scalar $X$ only couples to the up quark (i.e., not the charm or the top) and finally 
\begin{equation}
  {\cal L}_{\rm int}=- g(u_{R \alpha} d_{R \beta})X^{\alpha \beta} + \text{h.c.}
\end{equation}
where again the charged scalar only couples to the first generation quarks. Limits on the allowed two-dimensional parameter space for the coupling constant $g$ and mass $M_X$ would be very useful. 

Given   LHC, LEP, and flavor constraints our conclusions about the potential observability of $\Delta B=\pm \Delta L=-2$ processes in the laboratory are rather pessimistic. However, one of the non-minimal models we considered (Model VI) may give an observable rate for the  process $pp \rightarrow \mu^+\mu^+$ (at the nuclear level $(A,Z)\rightarrow (A-2, Z-2)+\mu^{+} \mu^{+}$) if our naive dimensional analysis underestimates the hadronic matrix element relevant for this process by more than an order of magnitude.


\begin{acknowledgments}
We thank Harvey Newman for help related to the LHC constraints on new colored scalars. 
This material is based upon work supported by the U.S. Department of Energy, Office of Science, Office of High Energy Physics, under Award Number DE-SC0011632 and by the Walter Burke Institute for Theoretical Physics.
\end{acknowledgments}

\bibliography{baryonBib}

\end{document}